\documentclass[prl,superscriptaddress,showpacs,twocolumn,epsf]{revtex4}
\usepackage{epsfig}
\usepackage[dvips]{color}

\def \zzz#1{{\framebox{#1}}}

\begin{document}


\title{Samaritan's Dilemma: Classical and quantum strategies in Welfare Game}
\author{\c{S}ahin Kaya \"Ozdemir}
\author{Junichi Shimamura}
\affiliation{CREST Research Team for Interacting Carrier
Electronics, The Graduate University for Advanced Studies
(SOKENDAI), Hayama, Kanagawa 240-0193, Japan }
\author{Fumiaki Morikoshi}
\affiliation{NTT Basic Research Laboratories, NTT Corporation, 3-1
Morinosato Wakamiya, Atsugi, Kanagawa 243-0198, Japan}
\author{Nobuyuki
Imoto} \affiliation{CREST Research Team for Interacting Carrier
Electronics, The Graduate University for Advanced Studies
(SOKENDAI), Hayama, Kanagawa 240-0193, Japan } \affiliation{NTT
Basic Research Laboratories, NTT Corporation, 3-1 Morinosato
Wakamiya, Atsugi, Kanagawa 243-0198, Japan}

\begin{abstract}
Effects of classical/quantum correlations and operations in game
theory are analyzed using Samaritan's Dilemma. We observe that
introducing either quantum or classical correlations to the game
results in the emergence of a unique or multiple Nash equilibria
(NE) which do not exist in the original classical game. It is
shown that the strategies creating the NE and the amount of
payoffs the players receive at these NE's depend on the type of
the correlation. We also discuss whether the Samaritan can resolve
the dilemma acting unilaterally.
\end{abstract}
\date{\today}
\pagestyle{plain} \pagenumbering{arabic} \maketitle

\section{I. Introduction}
Game theory is an interdisciplinary approach that can offer
insights into economic, political, and social situations which
involve decision makers who have different goals and preferences
and who know that their actions affect each other. Since the
activities of the decision makers include information processing
and a physical system is needed for the implementation of the
games, it is no surprise that, currently, game theory is becoming
an attractive research topic within the quantum information
community
\cite{Eisert1,Du,Flitney1,Eisert2,Meyer,Enk,Du2,Johnson3,Ben,Flitney2,Iqbal}.

By introducing quantum operations and/or quantum correlations into
the game theory, a number of games including Prisoner's Dilemma
(PD) \cite{Eisert1}, Battle of Sexes (BoS) \cite{Du}, Monty Hall
problem (MH) \cite{Flitney1}, Chicken Game (CG) \cite{Eisert2},
and matching pennies (MP) \cite{Meyer} have been studied.
Recently, there has been efforts to form a general theory of
quantum games \cite{Johnson1,Johnson2}. The results of the games
when played with the tools of quantum mechanics have been shown to
be very different from those of their classical counterparts, e.g.
the dilemma of the prisoners have been resolved in the quantum
game. Since the field is very new and on its way of development,
it is necessary to study quantum versions of various classical
games and analyze them within the paradigm of quantum mechanics to
understand the new features introduced by quantum mechanics
better.

A game $\Gamma$ can be denoted by $\Gamma=[N,(S_{i})_{i\in
N},(u_{i})_{i\in N}]$ where $N$ is the set of players, $S_{i}$ is
the strategy set for $i$-th player, and $u_{i}$ is the payoff
function from the set of all possible strategy combinations into
the set of real numbers for the $i$-th player. Then the payoff for
the $i$-th player can be denoted as $u_{i}(s)$ where $s$ is the
combination of the strategies implemented by all players. A
two-player game with each player having the strategy sets $S_{A}$
and $S_{B}$
\begin{table}[h]
\begin{center}
\begin{tabular}{ccc}
\hline
& Bob:$\alpha_{1}$~~~~ & Bob: $\beta_{1}$\\
\hline
 Alice: $\alpha_{1}$~~~~ & $(a,w)$~~~~&$(b,x)$   \\
 Alice: $\beta_{1}$~~~~ & $(c,y)$~~~~&$(d,z)$  \\
\hline
\end{tabular}
\caption{2-by-2 game payoff matrix\label{tab:tc}}
\end{center}
\end{table}
can be represented as $\Gamma=[\{{\rm
Alice,Bob}\},(S_{A},S_{B}),(u_{A},u_{B})]$ from which, for
example, Alice's payoff for the strategy combination of
$(\alpha,\beta)$ where $\alpha\in S_{A}$ and $\beta\in S_{B}$ can
be written as $u_{A}(\alpha,\beta)$. In classical game theory, any
game is fully described by its payoff matrix. Based on the nature
of payoffs, games can be classified in three different ways (i)
symmetric, $u_{A}(\alpha,\beta)=u_{B}(\beta,\alpha)$, and
asymmetric games, $u_{A}(\alpha,\beta)\neq u_{B}(\beta,\alpha)$,
(ii) zero-sum, $u_{A}(\alpha,\beta)+u_{B}(\alpha,\beta)=0$ for
$\forall \alpha\in S_{A}$, and non-zero sum games, if for $\exists
\alpha\in S_{A}$ and $\exists \beta\in S_{B}$,
 $u_{A}(\alpha,\beta)+u_{B}(\alpha,\beta)\neq0$ , and (iii)
coordination if the game has at least one Nash equilibrium (NE),
and discoordination games if there is no NE in pure strategies
\cite{Eric}.

Nash equilibrium is the most commonly used equilibrium concept for
strategic games, and it can be regarded as the steady state of the
strategic interaction. The concept of NE is based on the premises
that each player acts rationally according to the belief he/she
has on the other player, and that the beliefs of each player about
the other one is correct. Once the players acts according to NE,
no one can take another action to unilaterally deviate from it in
order to increase his/her payoff. In an NE, each player's choice
of action is the best response to the actions taken by the other
player. Although in pure strategies an NE need not exist, there is
always at least one NE in mixed strategies where the players are
allowed to randomize among their action. The most common
difficulty encountered in the concept of NE is that NE need not be
unique. There might be multiple NE's which avoids making sharp
decisions. In such cases, certain NE's can be isolated as ``focal"
in that they are clearly better for all players, and they yield a
higher payoff to every player than any other NE's. If there exists
only one NE like
\begin{table}[h]
\begin{center}
\begin{tabular}{cccc}
\hline & Bob: Work (W)&$~
$ & Bob: Loaf (L)\\
\hline
 Alice: Aid (A)& $(3,2)$&$\rightarrow$&$(-1,3)$   \\
 &$\uparrow$& &$\downarrow$\\
 \hspace{3mm}Alice: No Aid (N)&$(-1,1)$&$\leftarrow$&$(0,0)$  \\
\hline
\end{tabular}
\caption{Payoff matrix for Welfare Game where Alice is the
Samaritan and Bob is the beneficiary. Arrows show the best
response of a player for a chosen strategy of the other player.
Any strategy pointed by two arrows is an NE. In this game, there
is no such a situation, therefore the game has no NE in classical
pure strategies.\label{tab:tc1a}}
\end{center}
\end{table}
\noindent this, then players will be self-enforced to play it,
thus solving the difficulty induced by multiple NE's.

An example of payoff matrix is given in Table {\ref{tab:tc}} for a
two-player-two-strategy game where
$S_{A}=S_{B}=\{\alpha_{1},\beta_{1}\}$. The payoffs of players for
all possible strategy combinations are represented as $(..,..)$
with the first element being Alice's payoff and the second Bob's.
This payoff matrix represents a symmetric game if $a=w$,
$(b,x)=(y,c)$, and $d=z$; and it represents a zero-sum game if
$a+w=b+x=c+y=d+z=0$. When $a>c$,$~d>b$,$~x>w$,$~y>z$ or
$c>a$,$~b>d$,$~w>x$,$~z>y$ is satisfied, it represents a
discoordination game. On the other hand, for example, if
$a>c$,$~d>b$,$~w>x$,$~z>y$, the game becomes a coordination game
with two NE's in pure strategies.

If we look at the games which have been studied using the quantum
mechanical tools, we see that, except the MH and MP games, which
are sequential move games, all the games are simultaneous move
games. The properties of the simultaneous move games studied so
far, then can be listed as: PD\&CG: non-zero-sum, symmetric,
coordination;  and BoS: non-zero-sum, asymmetric, coordination.
Although, for these games, it has been shown that pure quantum
strategies bring interesting features and, in some cases, resolve
the dilemmas of the game, it is not very clear whether quantum
strategies can resolve problems in other types of games or payoff
matrices. It is, therefore, interesting to study other games
belonging to different classes and having different payoff matrix
structures. One of such payoff matrices is that of the so called
``Welfare game" or as often referred to as ``Samaritan's dilemma,"
which can be classified as a non-zero-sum, asymmetric, and
discoordination game. This game is chosen because it appears in a
wide range of social, economical, and political issues and the
dilemma present in the game is as strong as the dilemmas existing
in PD and BoS games.

In this paper, we study the Welfare game whose payoff matrix is
shown in Table \ref{tab:tc1a}. The paper is organized as follows:
In Sec. II, we will introduce the classical game and discuss the
game using both pure and mixed classical strategies without any
shared correlation between the players. Then in Sec. III, we will
introduce the quantum version of the game and show the features of
the game in different situations: (i) quantum operations with
shared quantum correlations, (ii) quantum operations with shared
classical correlations, (iii) classical operations with shared
classical correlations, and (iv) classical operations with shared
quantum correlations. Section IV will include a discussion of
whether the players can resolve the dilemma by unilateral actions,
and finally, in Sec. V, we will give a brief summary and
discussion of our results.


\section{II. Welfare game and Samaritan's dilemma}
The Samaritan's Dilemma arises whenever actual or anticipated
``altruistic" behavior of the ``Samaritan" (Alice) leads to
exploitation on the part of the potential beneficiary (Bob), such
that Alice suffers a welfare loss when compared to the situation
that would have been obtained if Bob had not acted strategically
\cite{Buch,Eric}. Most people have personally experienced this
dilemma when confronted with people ``in need." Although there is
a desire to help those people who cannot help themselves, there is
the recognition that a handout may be harmful to the long-run
interests of the recipient. If the condition of the person in need
is beyond that person's control, then there is no dilemma for
Samaritan. However, the person in need can influence or create
situations which will evoke Samaritan's help. Then, a dilemma
arises because Samaritan wants to help, however, the action of the
person in need leads an increase in the amount of help which is
not desirable for the Samaritan. Moreover, Samaritan cannot
retaliate to minimize or stop this exploitation because doing so,
which is a punishment for the people in need, will harm the
Samaritan's own interests in the short run.  The Samaritan's
Dilemma, which was modelled as a two-person strategic game by the
Nobel Laureate economist James Buchanan, is widespread in a wide
range of distinct issue areas from international politics,
government welfare programs, and family issues  \cite{Buch}.

The game and the payoff matrix studied in this paper are taken
from Ref.\cite{Eric} where the specific game is named as The
Welfare Game. In this game, Alice wishes to aid Bob if he searches
for work but not otherwise. On the other hand, Bob searches for
work if he cannot get aid from Alice.

Now let us analyze this classical game for pure and mixed
strategies. The strategy combinations can be listed as $(A,W)$,
$(A,L)$, $(N,W)$, and $(N,L)$.  When players use only pure
strategies, there is no dominant strategy for neither of the
players and, moreover, there is no NE, that is why this is a
discoordination game. This can be explained as follows: (A,W) is
not an NE because if Alice chooses A, Bob can respond with
strategy L where he gets a better payoff (three) as shown with
arrow in Table \ref{tab:tc1a}. (A,L) is not an NE because, in this
case, Alice will switch to N. The strategies (N,L) and (N,W) are
not NE either, because for the former one Bob will switch to W to
get payoff one, whereas for the latter case Alice will switch to A
to increase her payoff from $-1$ to $3$.  Therefore, this game has
no NE when played with pure classical strategies.

In mixed classical strategies, we assign the probabilities $p$ and
$(1-p)$ to the events that Alice chooses strategies A and N,
respectively. In the same way, for Bob's choices of W and L, we
assign the probabilities $q$ and $(1-q)$. Then the payoffs for
Alice and Bob become
\begin{eqnarray}\label{N01}
\$_{A}=3pq-p(1-q)-q(1-p),\nonumber \\
\$_{B}=2pq+3p(1-q)+q(1-p).
\end{eqnarray} It can be easily shown that $p=0.5$ and $q=0.2$ correspond to the NE for the game with average
payoffs given as $\$_{A}=-0.2$ and $\$_{B}=1.5$. In this case, the
payoff of Alice is negative which is not a desirable result for
her. (A,L) and (N,L) emerge as the most probable strategies with
probabilities $0.4$.

Samaritan's dilemma game is a very good example of how the
altruistic and selfish behavior of the players can affect a game
and its dynamics. It is different than the other already studied
games because it has no NE in pure strategies and it is only Alice
who is facing the dilemma. Games with no NE are interesting
because they represent situations in which individual players
might never settle down to a stable solution. Therefore, the
important step in resolving the dilemma is to find an NE on which
the players can settle down.

In this work, we look for the strategies utilizing classical or
quantum mechanical toolboxes to find solutions to the following
problem: Is there a unique NE where the players can settle? If the
answer to this question is YES, then we ask how the payoffs of the
players compare with each other and to what extent this NE
strategy can resolve the dilemma: CASE I: $\$_{A}<0$ (insufficient
solution), CASE II: $0\leq\$_{A}\leq\$_{B}$ (weak solution), and
CASE III: $0\leq\$_{B}<\$_{A}$ (strong solution). Finding a unique
NE is the first step in resolving the dilemma. The positive
payoffs players receive at a unique NE means that both players are
satisfied with the outcome and there is no loss of resources for
Alice (CASE II and III). The most desirable solution of the
dilemma for Alice is represented in CASE III which can easily be
seen from the original payoff matrix shown in Table
\ref{tab:tc1a}.

\section{III. Quantum version of the game and effects of various strategic spaces}
Since the Welfare Game is a two-player game, quantum strategies
can be introduced using the physical model
\begin{figure}[h]
\vspace{0mm} \hspace{0mm}\epsfxsize=8cm \epsfbox{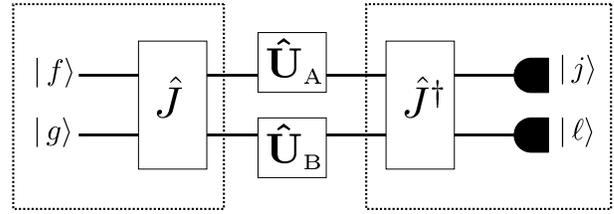}
\vspace{-2mm}\caption[]{Schematic configuration of the quantum
version of $2\times 2$ strategic games. The operations inside the
dotted boxes are performed by the referee. }\label{Fig1}
\end{figure}
\noindent given in Fig.\ref{Fig1} \cite{Eisert1}. In this physical
model, starting from an initial product state
$|\psi_{0}\rangle=|fg\rangle$, the referee prepares the maximally
entangled state (MES)
$\hat{\rho}_{\rm{in}}=\hat{J}|\psi_{0}\rangle\langle\psi_{0}|\hat{J}^{\dagger}$
where $\hat{J}$ is the entangling operator defined as
\begin{eqnarray}\label{N04}
\hat{J}|fg\rangle=\frac{1}{\sqrt{2}}[~|fg\rangle+i(-1)^{(f+g)}|(1-f)(1-g)\rangle~]
\end{eqnarray}\noindent with $f,g=0,1$. In this
scheme, $\hat{J}$ is chosen such that the original classical game
can be reproduced if the classical strategies A and W are defined
with the identity operator $\sigma_{0}$, and N and L are defined
with the bit flip operator $i\sigma_{y}$ when
$|\psi_{0}\rangle=|00\rangle$. Assignment of the operators to the
classical strategies denoted by A,W,N, and L is initial state
dependent, i.e., when $|\psi_{0}\rangle=|01\rangle$, A and L
should be defined with $\sigma_{0}$, while N and W with
$i\sigma_{y}$.

Let us assume that the referee prepares $\hat{\rho}_{\rm{in}}$ and
sends one of the qubits of this state to Alice and the other one
to Bob. Alice and Bob perform local operations, respectively,
denoted by $\hat{U}_{A}$ and $\hat{U}_{B}$ on each of their qubits
separately. After these operations, the resulting state becomes
\begin{equation}\label{N02}
\hat{\rho}_{\rm{out}}=(\hat{U}_{A}\otimes\hat{U}_{B})\hat{\rho}_{\rm{in}}(\hat{U}^{\dagger}_{A}\otimes\hat{U}^{\dagger}_{B}).
\end{equation}
The referee who receives this final state first performs
$\hat{J}^{\dagger}\hat{\rho}_{\rm{out}}\hat{J}$ and then makes a
projective measurement $\Pi_{n}=|j \ell\rangle\langle j
\ell|_{\{j,\ell={0,1}\}}$ with $n=2j+\ell$ corresponding to the
projection onto the orthonormal basis
$\{|\rm{AW}\rangle,|\rm{AL}\rangle,|\rm{NW}\rangle,|\rm{NL}\rangle\}=\{|00\rangle,|01\rangle,|10\rangle,|11\rangle\}$.
According to the measurement outcome $n$, the referee assigns to
each player the payoff chosen from the payoff matrix given in
Table \ref{tab:tc1a}, e.g., if $n=0$ then the payoffs assigned to
Alice and Bob are $3$ and $2$, respectively. Then the average
payoff of the players can be written as
\begin{eqnarray}\label{N03}
&&\$_{A}=\sum_{n} a_{n}\underbrace{\rm{Tr}(\Pi_{n}\hat{J}^{\dagger}\hat{\rho}_{\rm{out}}\hat{J})}_{P_{j \ell}}\nonumber\\
&&\$_{B}=\sum_{n}
b_{n}\underbrace{\rm{Tr}(\Pi_{n}\hat{J}^{\dagger}\hat{\rho}_{\rm{out}}\hat{J})}_{P_{j
\ell}}
\end{eqnarray} with $a_{\{n=0,1,2,3\}}=\{3,-1,-1,0\}$ and
$b_{\{n=0,1,2,3\}}=\{2,3,1,0\}$ being the payoffs of Alice and Bob
chosen from Table \ref{tab:tc1a}. ${P_{j \ell}}$ represents the
probability of obtaining the measurement outcome $n$.

Players are restricted to choose their operators, $\hat{U}_{A}$
and $\hat{U}_{B}$, from two- and one-parameter SU(2) as
\begin{eqnarray}\label{N05}
\hat{U}_{A}=\left(%
\begin{array}{ccc}
  e^{i\phi_{A}}\cos\frac{\theta_{A}}{2} && \sin\frac{\theta_{A}}{2}\\
&\\
  -\sin\frac{\theta_{A}}{2} && e^{-i\phi_{A}}\cos\frac{\theta_{A}}{2} \\
\end{array}%
\right),
\end{eqnarray}\noindent with $0\leq\phi_{A}\leq\pi/2$ and $0\leq\theta_{A}\leq\pi$. Bob's operator, $\hat{U}_{B}(\theta_{B},\phi_{B})$, can be
obtained in the same way by replacing $A$ with $B$ in
Eq.(\ref{N05}) \cite{Text1}. In the following subsections, we
investigate the evolution of the Welfare Game with various
strategy sets: (A) Quantum operations and quantum correlations,
(B) quantum operations and classical correlations, (C) classical
operations and classical correlations, and (D) classical
operations and quantum correlations.

\subsection{A. Quantum operations and quantum correlations}
Before proceeding further, it is worth noting that if the players
use quantum operations with no shared correlated states, then the
situation is similar to playing the game with classical mixed
strategies which is discussed in Sec. II.
\subsubsection{1. One-parameter SU(2) operators}

This operator set for Alice and Bob can be obtained from
Eq.(\ref{N05}) by setting $\phi_{A}=0$ and $\phi_{B}=0$. First,
let us assume that $|\psi_{0}\rangle=|00\rangle$, thus the
entangled state becomes
$\hat{\rho}_{\rm{in}}=|\psi_{\rm{in}}\rangle\langle\psi_{\rm{in}}|$
with $|\psi_{\rm{in}}\rangle=[|00\rangle+i|11\rangle]/\sqrt{2}$.
Then the payoff for Alice and Bob are found as
\begin{eqnarray}\label{N06}
\$_{A}&=&\frac{1}{4}[1+3(\cos \theta_{A}+\cos\theta_{B})+5\cos\theta_{A}\cos\theta_{B}],\nonumber \\
\$_{B}&=&\frac{1}{2}[3+2\cos
\theta_{A}-\cos\theta_{A}\cos\theta_{B}].
\end{eqnarray}\noindent It is seen from Eq.(\ref{N06}) that
$\$_{B}$ is always positive whereas $\$_{A}$ is either positive or
negative depending on the action of Bob. Therefore, Alice cannot
always get a positive payoff and hence cannot resolve her dilemma
by acting unilaterally and independently of Bob's strategy. After
some straightforward calculations, it can be found that players
can achieve an NE if they choose
$(\theta_{A}=\pi/2,\cos\theta_{B}=-3/5)$ corresponding to the
operators $\hat{U}_{A}=(\sigma_{0}+i\sigma_{y})/ \sqrt{2}$ and
$\hat{U}_{B}=(\sigma_{0}+i2\sigma_{y})/ \sqrt{5}$ where
$\sigma_{i}$ are the Pauli operators. At this unique NE, the
payoffs for the players become $\$_{A}=-0.2$ and $\$_{B}=1.5$.

On the other hand, when the initial state is
$|\psi_{0}\rangle=|01\rangle$, an NE is found at
$(\theta_{A}=\pi/2,\cos\theta_{B}=3/5)$ corresponding to
$\hat{U}_{A}=(\sigma_{0}+i\sigma_{y})/ \sqrt{2}$ and
$\hat{U}_{B}=(2\sigma_{0}+i\sigma_{y})/ \sqrt{5}$ with the payoffs
$\$_{A}=-0.2$ and $\$_{B}=1.5$.

It is seen that playing the quantum version of the game with
one-parameter set of operators and a shared MES between the
players reproduces the results of the classical mixed strategy.
Both cases have one NE with the same payoffs $\$_{A}=-0.2$ and
$\$_{B}=1.5$. This strategy set resolves the dilemma, but with an
insufficient solution for Alice (CASE I).

\subsubsection{2. Two-parameter SU(2) operators}
This set of quantum operations is defined as given in
Eq.(\ref{N05}). For the case when $|\psi_{0}\rangle=|00\rangle$,
the expressions for the payoffs are given as
$\$_{A}=3P_{00}-P_{01}-P_{10}$ and $\$_{B}=2P_{00}+3P_{01}+P_{10}$
with
\begin{eqnarray}\label{N08}
P_{00}&=&\cos^{2}(\theta_{A}/2)\cos^{2}(\theta_{B}/2)\cos^{2}(\phi_{A}+\phi_{B})\nonumber\\
P_{01}&=&|x\sin\phi_{B}-y\cos\phi_{A}|^{2}\nonumber\\
P_{10}&=&|x\cos\phi_{B}-y\sin\phi_{A}|^{2}
\end{eqnarray} where $P_{j \ell}$ is calculated from Eq. (\ref{N03}) and we
set $x=\sin(\theta_{A}/2)\cos(\theta_{B}/2)$ and
$y=\cos(\theta_{A}/2)\sin(\theta_{B}/2)$. A straightforward
analysis of these equations reveals that there is a unique NE
which appears at $(\theta_{A}=0,\phi_{A}=\pi/2,
\theta_{B}=0,\phi_{B}=\pi/2)$ corresponding to
$\hat{U}_{A}=\hat{U}_{B}=i\sigma_{z}$ with the payoff
$(\$_{A},\$_{B})$ given as $(3,2)$. We see that introducing
quantum operations and correlations results in the emergence of
this NE point which cannot be seen when the game is played with
pure classical strategies. Thus the original discoordination game
becomes a coordination game. This unique NE gives Alice the
highest payoff she can get in this game. This is the NE point
where Alice always wants to achieve because both players benefit
from playing the game, moreover, Alice does not lose her
resources. Therefore, Alice's dilemma is resolved in the stronger
sense (CASE III). It is also seen that players receive higher
payoffs than those obtained if a classical mixed strategy were
used.

This new move of the players can be included into the game payoff
matrix as a new strategy \cite{Enk}. This, indeed, shows that the
quantum strategy introduced into the game transformed the original
($2\times2$) game into a new game which can be described with the
new ($3\times3$) payoff matrix. The new payoff matrix of the game
is shown in Table \ref{tab:tf2} where the new strategy
$\hat{U}_{A}=\hat{U}_{B}=i\sigma_{z}$ for the players are depicted
as M.
\begin{table}[h]
\begin{center}
\begin{tabular}{cccccc}
\hline & Bob: W&~& Bob: L&~& Bob:M\\
\hline
 Alice: A&\hspace{1.2mm} \colorbox[rgb]{0.82,0.82,0.82}{$(3,2)~~~~~$}&&\hspace{-4mm}\colorbox[rgb]{0.82,0.82,0.82}{$~~~~~~(-1,3)$}&&$(0,0)$   \\
 Alice: N&\hspace{1.3mm}  \colorbox[rgb]{0.82,0.82,0.82}{$(-1,1)~~~$}&&\hspace{-4mm}\colorbox[rgb]{0.82,0.82,0.82}{$~~~~~~~~(0,0)$}&&$(-1,3)$   \\
\hspace{0mm}Alice: M& $(0,0)$&&$(-1,1)$&&$\zzz{(3,2)}$   \\
\hline
\end{tabular}
\caption{The new payoff matrix for the Welfare game which includes
the new strategy for the players to resolve the dilemma of the
game. The new strategy $i\sigma_{z}$ which is incorporated into
the payoff matrix as the classical strategy M is the operator
which resolves the dilemma of the game when players share the
quantum correlation $[|00\rangle+i|11\rangle]/\sqrt{2}$, and use
two-parameter SU(2) operators. The original classical game payoff
matrix (gray colored) is seen as a $2\times2$ sub-matrix. The
entry of the matrix in solid box corresponds to the NE.
\label{tab:tf2}}
\end{center}
\end{table}
This new game payoff matrix includes the original classical game
payoff matrix as its subset. Consequently, one can say that the
classical game is a subgame of its quantum version. In other
words, one can say that in order to reproduce the same results of
this quantum game in classical settings, the players, Alice and
Bob, should be given the strategy sets $\{{\rm A, N, M }\}$ and
$\{{\rm W, L, M }\}$, respectively. Then in order to classically
communicate their chosen strategy to the referee, each player
needs 2 classical bits (c-bits) resulting in a total of 4 c-bits.
Note that the same task is completed using four qubits (two
prepared by the referee and distributed, and two qubits sent by
the players back to the referee after being operated on) when
shared entanglement is used. This corresponds to a communication
cost of 2 e-bits (1 e-bit for the referee and 1 e-bit for the
players).

On the other hand, when $|\psi_{0}\rangle=|01\rangle$, we found
four NE's with equal payoffs $(\$_{A},\$_{B})=(3,2)$ when the
players choose the following quantum operations
$(\theta_{A}=\pi,\phi_{A}=0, \theta_{B}=0,\phi_{B}=\pi/2)$,
$(\theta_{A}=\pi/2,\phi_{A}=0,\theta_{B}=\pi/2,\phi_{B}=\pi/2)$,
$(\theta_{A}=2\pi/3,\phi_{A}=0,\theta_{B}=\pi/3,\phi_{B}=\pi/2)$,
and
$(\theta_{A}=3\pi/4,\phi_{A}=0,\theta_{B}=\pi/4,\phi_{B}=\pi/2)$.
These, respectively, correspond to the operators
$(\hat{U}_{A},\hat{U}_{B})$ as follows $({\rm
N,P})=(i\sigma_{y},i\sigma_{z})$, (${\rm
T,Q})=\frac{(\sigma_{0}+i\sigma_{y})}{\sqrt{2}},\frac{i(\sigma_{z}+\sigma_{y})}{\sqrt{2}})$,
$({\rm
Y,R})=\frac{\sigma_{0}+i\sqrt{3}\sigma_{y}}{2},\frac{i(\sqrt{3}\sigma_{z}+\sigma_{y})}{2})$,
and $({\rm
Z,S})=(\gamma_{0}(\sigma_{0}+ia_{1}\sigma_{y}),ib_{0}(\sigma_{z}+b_{1}\sigma_{y}))$
where $\gamma_{0}=\cos(3\pi/8)$, $\gamma_{1}=\tan(3\pi/8)$,
$\delta_{0}=\cos(\pi/8)$, and $\delta_{1}=\tan(\pi/8)$. These NE's
have higher payoffs for both players than those obtained when a
classical mixed strategy is used. However, in this pure quantum
strategy case, the dilemma of the Samaritan (Alice) still
continues, because both players cannot decide which NE point to
choose. For example, if Alice thinks that Bob will play
$i\sigma_{z}$ then she will play $i\sigma_{y}$ to reach at the
first NE. However, since this is a simultaneous move game and
there is no classical communication between the players, Bob may
play $\frac{i(\sigma_{z}+\sigma_{y})}{\sqrt{2}}$ (because this
action will take him to the second NE point) while Alice plays
$i\sigma_{y}$. Such a case will result in the case $\$_{A}<\$_{B}$
and will lower the payoffs of both players. Therefore, still a
dilemma exists in the game; however the nature of dilemma has
changed.

As we have done for the previous case, these four new strategies
which give NE's can be added to the payoff matrix of the original
game. This results in a $(5\times6)$ payoff matrix, as depicted in
Table \ref{tab:tf245}, which has the original classical game as a
subgame. If the players are given this payoff matrix and
restricted to classical communication then a total of 6 c-bits are
needed to play this game whereas in quantum strategies what is
needed is 2 e-bits.

\begingroup
\squeezetable
\begin{table*}
\begin{tabular}{cccccccccccc}
\hline &Bob: W&~&Bob: L&~&Bob: P&~&Bob: Q&~&Bob: R&~&Bob: S\\
\hline
Alice: A&\colorbox[rgb]{0.82,0.82,0.82}{~~~~~$(3,2)$~~~~~~}&&\hspace{-10mm}\colorbox[rgb]{0.82,0.82,0.82}{$~~~~~~~~~(-1,3)$}&&$(-1,1)$&& $(1,\frac{3}{2})$&&$(0,\frac{5}{4})$&&$(1-\sqrt{2},\frac{6-\sqrt{2}}{4})$   \\
Alice: N&\colorbox[rgb]{0.82,0.82,0.82}{\hspace{0.3mm}~~~$(-1,1)~~~~~~$}&&\hspace{-10mm}\colorbox[rgb]{0.82,0.82,0.82}{$~~~~~~~~~~~~(0,0)$}&&$\zzz{(3,2)}$&& $(1,\frac{3}{2})$&&$(2,\frac{7}{4})$&&$(1+\sqrt{2},\frac{6+\sqrt{2}}{4})$   \\
Alice: T& $(1,\frac{3}{2})$&&$(-\frac{1}{2},\frac{3}{2})$&&$(1,\frac{3}{2})$&& $\zzz{(3,2)}$&&$(1+\sqrt{3},\frac{6+\sqrt{3}}{4})$&&$(1+\sqrt{2},\frac{6+\sqrt{2}}{4})$   \\
Alice: Y& $(0,\frac{5}{4})$&&$(-\frac{1}{4},\frac{3}{4})$&&$(2,\frac{7}{4})$&& $(1+\sqrt{3},\frac{6+\sqrt{3}}{4})$&&$\zzz{(3,2)}$&&$(1+\sqrt{2+\sqrt{3}},\frac{6+\sqrt{2+\sqrt{3}}}{4})$   \\
Alice: Z& $(1-\sqrt{2},\frac{6-\sqrt{2}}{4})$&&$(\frac{\sqrt{2}-2}{4},\frac{3(2-\sqrt{2})}{4})$&&$(1+\sqrt{2},\frac{6+\sqrt{2}}{4})$&& $(1+\sqrt{2},\frac{6+\sqrt{2}}{4})$&&$(1+\sqrt{2+\sqrt{3}},\frac{6+\sqrt{2+\sqrt{3}}}{4})$&&$\zzz{(3,2)}$   \\
 \hline
\end{tabular}
\caption{ New payoff matrix for the Welfare game when players
share the quantum correlation $[|01\rangle-i|10\rangle]/\sqrt{2}$,
and use two-parameter SU(2) operators. The operators which results
in new NE's are included together with the original payoff matrix
(upper left $2\times2$ sub-matrix). The elements in square boxes
correspond to NE's in this game. Gray colored $(2\times2)$
submatrix corresponds to the original classical game payoff
matrix. \label{tab:tf245}}
\end{table*}
\endgroup
\begin{table}[h]
\begin{center}
\begin{tabular}{cccc}
\hline & $~~~~(\hat{U}_{A},\hat{U}_{B})$&$~
$ &$~~~(\$_{A},\$_{B})$ \\
\hline
$p=1/4$& $~~~~(\sigma_{0},~\sigma_{0})$&$~$&$~~~(0,~11/4)$   \\
$~$& $~~~~(i\sigma_{y},~i\sigma_{z})$&$~$&$~~~(2,~9/4)$   \\
$p=1/2$& $~~~~(\sigma_{0},~i\sigma_{y})$&$~$&$~~~(1,~5/2)$   \\
$~$& $~~~~(i\sigma_{y},~i\sigma_{z})$&$~$&$~~~(1,~5/2)$   \\
$~$& $~~~~(i\sigma_{z},~i\sigma_{z})$&$~$&$~~~(1,~5/2)$   \\
$p=3/4$& $~~~~(\sigma_{0},~i\sigma_{y})$&$~$&$~~~(0,~11/4)$   \\
$~$& $~~~~(i\sigma_{z},~i\sigma_{z})$&$~$&$~~~(2,~9/4)$   \\
\hline
\end{tabular}
\caption{Strategies of players at the NE points and their
corresponding payoffs when the source is corrupted. The players
should know the characteristic of the source to decide on their
strategy. $p$ and $(1-p)$ are, respectively, the probability that
$|00\rangle\langle00|$ and $|01\rangle\langle01|$ are sent from
the source. \label{tab:prob}}
\end{center}
\end{table}
With these results, it is seen that the dynamics of the game, that
is whether the dilemma is resolved or not, the strategies, and the
payoffs of the players depend on the initial state and hence on
the shared MES. So far, we have considered the case that the
referee not only prepares the MES but also informs the players of
the initial state. Thus the players knowing the shared MES can
choose their best moves to resolve the dilemma. If they do not
know which type of the MES they share, the players cannot decide
which of the above discussed strategies to apply. Then it becomes
interesting to ask the question of what the best strategy is for
the players if they do not know the input state or even more
interesting is the question of what they can do if the source the
referee uses is corrupt.

Assuming that the source prepares the state $|00\rangle$ with
probability $p$ and $|01\rangle$ with probability $1-p$, the state
distributed to the players by the referee becomes
$\hat{\rho}_{\rm{in}}=p\hat{J}|00\rangle\langle00|\hat{J}^{\dagger}+(1-p)\hat{J}|01\rangle\langle01|\hat{J}^{\dagger}$.
Hence the payoffs for the players become as
$\$_{A}=p\$_{A}^{'}+(1-p)\$_{A}^{''}$ and
$\$_{B}=p\$_{B}^{'}+(1-p)\$_{B}^{''}$ where the expressions with
$'$ and $''$ corresponds to the payoffs the players receive when
the input state is $|00\rangle$ and $|01\rangle$, respectively.
This modification of the payoffs results in the emergence of new
strategies and NE's for different values of $p$. The problem
becomes very unpleasant for the players because the number of NE's
increases preventing the players from resolving the dilemma. To
give an idea on the strategies of the players which give rise new
NE's, we listed some of them in Table \ref{tab:prob}.

\subsection{B. Quantum operations and classical correlations}
\begin{figure}[h]
\vspace{0mm} \hspace{0mm}\epsfxsize=6cm
\vspace{5mm}\epsfbox{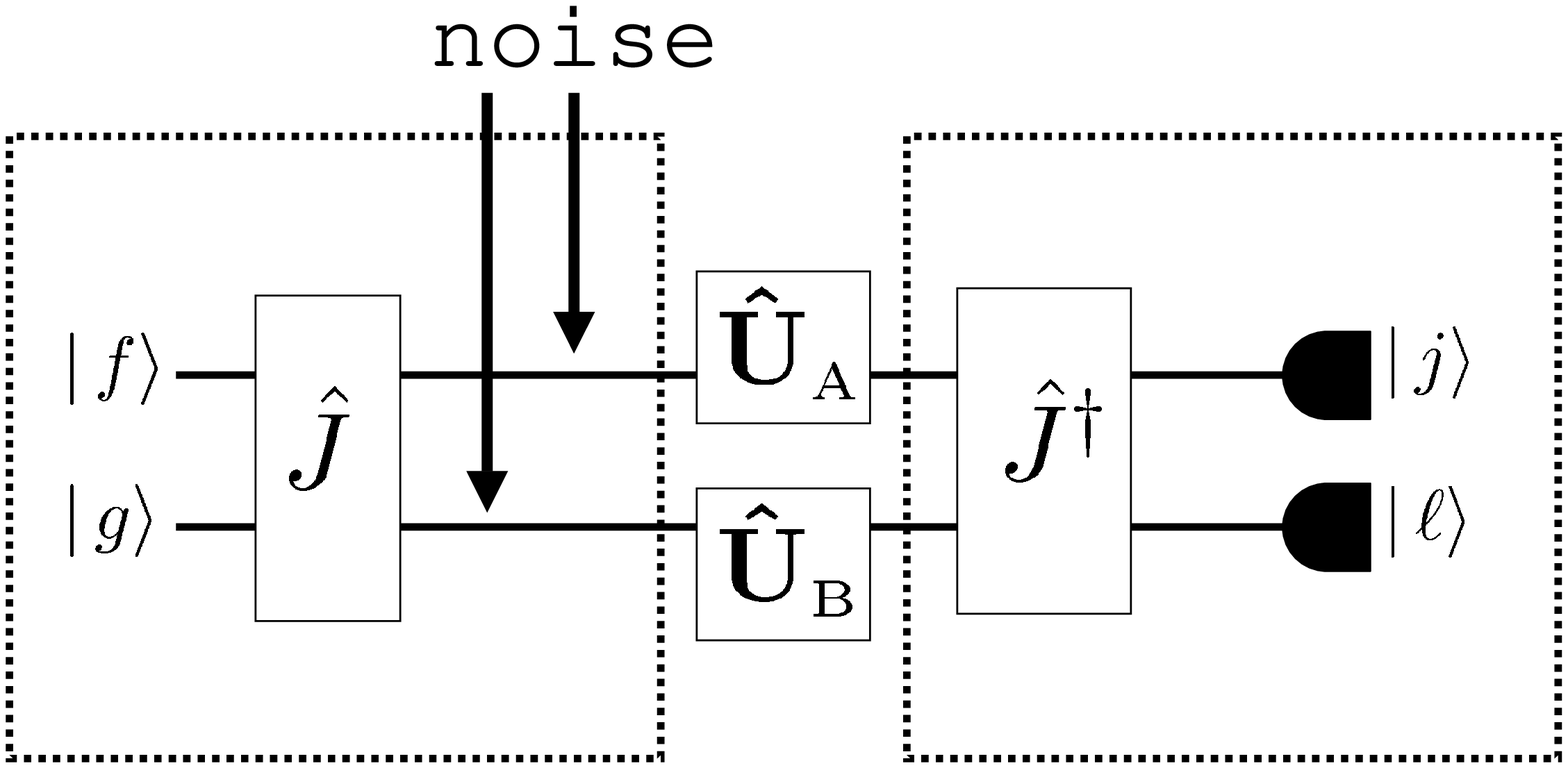} \epsfxsize=5cm
\epsfbox{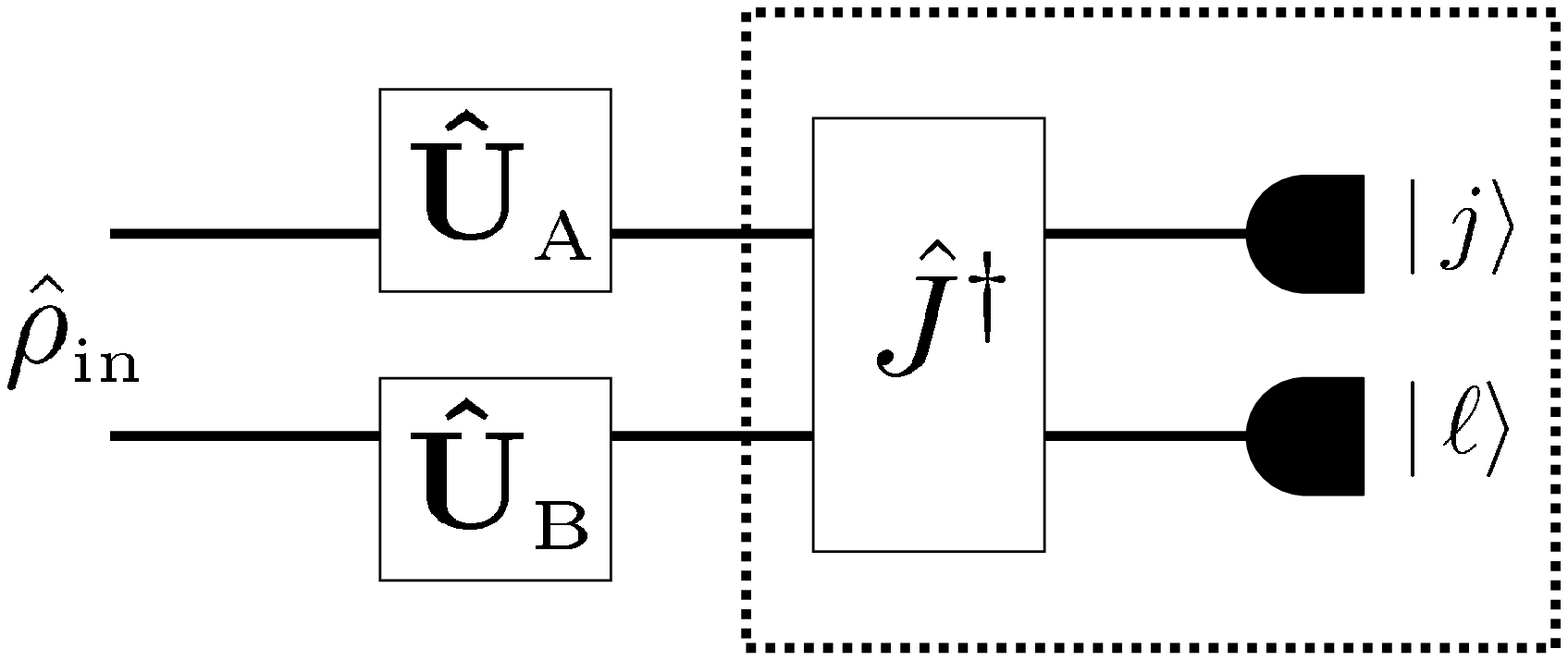}\vspace{-2mm}\caption[]{Introducing classical
correlations into the quantum version of the Welfare game.
$\hat{\rho}_{\rm in}$ is the classical correlations obtained from
the MES after it is subjected to noise. The operations inside the
dotted boxes are performed by the referee. }\label{Fig2}
\end{figure}

First let us assume that operators are chosen from the two
parameter SU(2) set while sharing classical correlations. This can
occur, for example, when one or both players induce phase damping
on the originally shared MES until the off-diagonal elements of
the density operator disappears or the referee distributes such a
state to them. If we assume that the initially shared MES is
$|\psi_{\rm{in}}\rangle=[|00\rangle-i|11\rangle]/\sqrt{2}$, then
the shared correlation will become the classical correlation
$\hat{\rho}_{\rm{in}}=[|00\rangle\langle00|+|11\rangle\langle
11|]/2$ after the damping. This new setting of the problem is
depicted in Fig.(\ref{Fig2}) from which the payoffs for the
players are calculated as
\begin{eqnarray}\label{N11}
\hspace{-1mm}\$_{A}\hspace{-2mm}&=&\hspace{-2mm}\frac{1}{4}[1+5\cos \theta_{A}\cos\theta_{B}-3\sin \theta_{A}\sin\theta_{B}\sin(\phi_{A}+\phi_{B})]\nonumber \\
\hspace{-1mm}\$_{B}\hspace{-2mm}&=&\hspace{-2mm}\frac{1}{2}[3-\cos
\theta_{A}\cos\theta_{B}-2\sin
\theta_{A}\sin\theta_{B}\cos\phi_{A}\sin\phi_{B}]
\end{eqnarray}\noindent Alice cannot obtain a positive
payoff by acting unilaterally and should try for NE. We find that
there appears an NE with a payoff $(0.25,1.5)$ when
$\hat{U}_{B}=(\sigma_{0}+i\sigma_{y})/ \sqrt{2}$ and Alice
restricts herself to one parameter SU(2) operators that is
$\phi_{A}=0$ and $\{\forall\theta_{A}: 0\leq\theta_{A}\leq\pi\}$.
Since neither of the players can make her/his payoff arbitrarily

\noindent high and positive independent of the action of the other
player, these operation sets for Alice and Bob which results in an
NE is the only self-enforcing solution. Considering that the
players are rational they will choose this set of operations,
consequently the dilemma for Alice will be resolved (CASE II).

When the shared correlation is in the form
$\hat{\rho}_{\rm{in}}=[|01\rangle\langle01|+|10\rangle\langle
10|]/2$, the payoffs of the players are given as in Eq.
(\ref{N11}) with all the $-$ being replaced by $+$ and vice verse
except the $+$ in the $\sin(..)$ term. Consequently, we find an NE
when $\hat{U}_{A}=(\sigma_{0}+i\sigma_{y})/ \sqrt{2}$ and
$\hat{U}_{B}=i(\sigma_{y}+\sigma_{z})/ \sqrt{2}$ are chosen which
gives the payoffs as $(\$_{A},\$_{B})=(2.5,1)$. This NE is a
self-enforcing point because by the choice of this specific
$\hat{U}_{A}$, Alice assures a positive payoff for herself and at
the same time for Bob which is independent of Bob's action. When
Alice applies this operation then the only thing that one can
expect from a rational Bob is to maximize his payoff by applying
the above $\hat{U}_{B}$. Therefore dilemma is resolved (CASE III).

If the players share a full rank classical correlation that is
$\hat{\rho}_{\rm{in}}=[|00\rangle\langle00|+|10\rangle\langle
10|+|01\rangle\langle01|+|11\rangle\langle 11|]/4$, they get
constant payoff $(0.25, 1.5)$ independent of their quantum
operations, and thus arriving at CASE II.

On the other hand, when the players are restricted to
one-parameter SU(2) operators, they get the payoff $(0.25,1.5)$ at
the NE with $\hat{U}_{A}=\hat{U}_{B}=(\sigma_{0}+i\sigma_{y})/
\sqrt{2}$ independent of the classical correlation they share.

These results suggest that after the maximally entangled state is
distributed by the referee, Alice and/or Bob can induce damping to
her/his qubit so that the quantum correlation becomes a classical
one. Then they can have unique strategy to go to an NE where both
gets positive payoffs. It must be noted that these solutions of
the game cannot be achieved by unilateral move of Alice or Bob.
That is, the dilemma is resolved only with the condition that
players are enforced to play the strategy that will take both of
them to the NE.

\subsection{C. Classical operations and classical correlations}
Classical operations are a subset of quantum operations. The
players have two operations either the identity operator
$\sigma_{0}$ or the flip operator $i\sigma_{y}$. Here we consider
the classical correlations as in the above subsection.  Let us
assume that the classical correlation is
$\hat{\rho}_{\rm{in}}=[|00\rangle\langle00|+|11\rangle\langle
11|]/2$. Due to this classical correlation, the diagonal and
off-diagonal elements of the payoff matrix of the original game
(Table \ref{tab:tc1a}) are averaged out separately, resulting in a
new game with the
\begin{table}[h]
\begin{center}
\begin{tabular}{cccc}
\hline & Bob: $\sigma_{0}$&$~
$ & Bob: $i\sigma_{y}$ \\
\hline
 Alice: $\sigma_{0}$& $(1.5,1)$&$\rightarrow$&$(-1,2)$   \\
 &$\uparrow$& &$\downarrow$\\
 \hspace{3mm}Alice: $i\sigma_{y}$&$(-1,2)$&$\leftarrow$&$(1.5,1)$  \\
\hline
\end{tabular}
\caption{The new payoff matrix for the Welfare game when players
share the classical correlation
$\hat{\rho}_{\rm{in}}=[|00\rangle\langle00|+|11\rangle\langle
11|]/2$, and use classical operations. \label{tab:tc2}}
\end{center}
\end{table}

\noindent payoff matrix given in Table \ref{tab:tc2} without any
correlation. From Table \ref{tab:tc2}, it is seen that the game is
still a discoordination game with no NE and the dilemma survives.
When the classical correlation is
$\hat{\rho}_{\rm{in}}=[|01\rangle\langle01|+|10\rangle\langle
10|]/2$, the payoff matrix has the same properties as in Table
\ref{tab:tc2} with diagonal and off-diagonal elements
interchanged.

The game becomes more interesting if we now let the players choose
a mixed strategy when applying the classical operations, that is
Alice and Bob apply $\sigma_{0}$, respectively, with probabilities
$p$ and $q$, and $i\sigma_{y}$, respectively, with probabilities
$1-p$ and $1-q$. In this case, it can be easily shown that there
is a unique NE when $p=0.5$ and $q=0.2$ with payoff (0.25,1.5).
Although, the values of $p$ and $q$ at NE is the same as the case
discussed in Section II where only classical mixed strategy for
the operators are allowed without any correlated shared state,
there is an increase in the payoff of Alice from $-0.2$ to $0.25$
owing to the shared classical correlation. This is interesting
because an NE appears where both Alice and Bob have positive
payoff, implying that the dilemma is resolved in the weaker sense
(CASE II).

\subsection{D. Classical operations and quantum correlations}
Since in this quantization scheme, the classical operators of the
players $\sigma_{0}$ and $i\sigma_{y}$ commute with
$\hat{J}^{\dag}$, the operations in the physical system reduces to
players' application of their choice of classical operations
(identity or flip operator) onto the $|\psi_{0}\rangle$ from which
the referee generates the maximally entangled state according to
Eq. (\ref{N04}) and then referee's projective measurement onto the
four basis $\{|00\rangle,|01\rangle,|10\rangle,|11\rangle\}$ to
calculate their payoff according to the outcome. Then the payoff
matrix of the game becomes the same as that of the original one in
Table \ref{tab:tc1a}. The game is still a discoordination game
with no NE at pure strategies. Therefore, the characteristics of
the game when played with pure and/or mixed strategies is the same
as explained in Sec. II.

\section{IV. Bob restricted to only classical operations}

Since the dilemma in this game is that of Alice, it is natural to
ask the question whether she can escape from this dilemma by
restricting Bob to only classical operations while she uses
quantum operations. We assume that the players share the MES,
$[|00\rangle+i|11\rangle]/\sqrt{2}$, Alice uses a fixed quantum
operator $\hat{U}_{A}$ chosen from the set of general SU(2)
operators given as
\begin{eqnarray}\label{N05555}
\hat{U}_{A}=\left(%
\begin{array}{ccc}
  e^{i\phi_{A}}\cos\frac{\theta_{A}}{2} && e^{i\varphi_{A}}\sin\frac{\theta_{A}}{2}\\
&\\
  -e^{-i\varphi_{A}}\sin\frac{\theta_{A}}{2} && e^{-i\phi_{A}}\cos\frac{\theta_{A}}{2} \\
\end{array}%
\right),
\end{eqnarray}\noindent with $0\leq\phi_{A},\varphi_{A}\leq\pi/2$ and $0\leq\theta_{A}\leq\pi$,
and Bob applies a mixed classical strategy that is
he applies $\hat{U}_{B}^{0}=\sigma_{0}$ with probability $p_{0}$
and $\hat{U}_{B}^{1}=i\sigma_{y}$ with probability $p_{1}$ where
$p_{0}+p_{1}=1$. Then what is the best strategy for Alice to solve
her dilemma?

In this setting, the state after Alice and Bob apply their
operators becomes
\begin{eqnarray}\label{N13}
\hat{\rho}_{\rm{out}}&=&\sum_{k=0}^{1}p_{k}(\hat{U}_{A}\otimes\hat{U}^{k}_{B})\hat{\rho}_{\rm{in}}(\hat{U}^{\dagger}_{A}\otimes\hat{U}^{k\dagger}_{B})\nonumber\\
&=&p~\hat{\rho}_{\rm{out}}^{\prime}+(1-p)~\hat{\rho}_{\rm{out}}^{\prime\prime},
\end{eqnarray} where $\hat{\rho}_{\rm{out}}^{\prime}$ and $\hat{\rho}_{\rm{out}}^{\prime\prime}$ are the output density operators
when Bob applies $\sigma_{0}$ and $i\sigma_{y}$, respectively, and
we have

\begin{table}[h]
\begin{center}
\begin{tabular}{ccc}
\hline Condition& Strategy& Payoff\\
\hline
&&\\
 \hspace{-7mm}{\bf No Corr.}& $\left(\frac{\sigma_{0}+i\sigma_{y}}{\sqrt{2}},\frac{\sigma_{0}+i2\sigma_{y}}{\sqrt{5}}\right)$&$(-\frac{1}{5},\frac{3}{2})$ \\
{\bf Quant. Corr.} & &\\
\\$\frac{|00\rangle+i|11\rangle}{\sqrt{2}}$& $\left(\frac{\sigma_{0}+i\sigma_{y}}{\sqrt{2}},\frac{\sigma_{0}+i2\sigma_{y}}{\sqrt{5}}\right)^{\ddagger}$&$(-\frac{1}{5},\frac{3}{2})$   \\

\\
& $\left(i\sigma_{z},i\sigma_{z}\right)$&$(3,2)$   \\
 &&\\
$\frac{|01\rangle-i|10\rangle}{\sqrt{2}}$& $\left(\frac{\sigma_{0}+i\sigma_{y}}{\sqrt{2}},\frac{2\sigma_{0}+i\sigma_{y}}{\sqrt{5}}\right)^{\ddagger}$&$(-\frac{1}{5},\frac{3}{2})$   \\
\\
& $\left(i\sigma_{y},i\sigma_{z}\right)$&$(3,2)$   \\
 & $\left(\frac{\sigma_{0}+i\sigma_{y}}{\sqrt{2}},\frac{i\sigma_{z}+i\sigma_{y}}{\sqrt{2}}\right)$&$(3,2)$  \\
& $\left(\frac{\sigma_{0}+i\sqrt{3}\sigma_{y}}{2},\frac{i\sqrt{3}\sigma_{z}+i\sigma_{y}}{2}\right)$&$(3,2)$  \\
&$\left(\gamma_{0}(\sigma_{0}+i\gamma_{1}\sigma_{y}),i\delta_{0}(\sigma_{z}+\delta_{1}\sigma_{y})\right)$&$(3,2)$  \\
 \\
{\bf Class. Corr.} & &\\
 \\$~~~~~\frac{|00\rangle\langle00|+|11\rangle\langle11|}{2}$& $\left(\frac{\sigma_{0}+i\sigma_{y}}{\sqrt{2}},\frac{\sigma_{0}+i\sigma_{y}}{\sqrt{2}}\right)^{\ddagger}$&$(\frac{1}{4},\frac{3}{2})$\\
 &&\\
 & $\left(\phi_{A}=0,\frac{\sigma_{0}+i\sigma_{y}}{\sqrt{2}}\right)$&$(\frac{1}{4},\frac{3}{2})$\\
 &&\\
 \\
$~~~~~\frac{|01\rangle\langle01|+|10\rangle\langle10|}{2}$& $\left(\frac{\sigma_{0}+i\sigma_{y}}{\sqrt{2}},\frac{\sigma_{0}+i\sigma_{y}}{\sqrt{2}}\right)^{\ddagger}$&$(\frac{1}{4},\frac{3}{2})$\\
 &&\\
& $\left(\frac{\sigma_{0}+i\sigma_{y}}{\sqrt{2}},\frac{i(\sigma_{y}+\sigma_{z})}{\sqrt{2}}\right)$&$(\frac{5}{2},1)$\\
 &&\\
{\it Full rank} &$\left(\forall~ \hat{U}_{A}, \forall~ \hat{U}_{B}\right)$&$(\frac{1}{4},\frac{3}{2})$  \\
\\
 \hline
\end{tabular}
\caption{Effect of the quantum operations (one-or two-parameter
SU(2) operators) of the players when they share quantum or
classical correlations. Only the strategies
$(\hat{U}_{A},\hat{U}_{B})$ where NE points emerge and the payoffs
$(\$_{A},\$_{B})$ that the players get at the specified NE are
listed. $\sigma_{i}$'s denote the Pauli spin operators,
$\gamma_{0}$ and $\gamma_{1}$, respectively, are $\cos(3\pi/8)$
and $\tan(3\pi/8)$ while $\delta_{0}$ and $\delta_{1}$, are
$\cos(\pi/8)$ and $\tan(\pi/8)$, respectively. $^{\ddagger}$
denotes the results when players use one-parameter SU(2)
operators. Results are the same for one- and two-parameter SU(2)
operators for the cases of no correlation and when the shared
correlated state is full rank.  Classical and quantum correlations
are denoted by ``Class. Corr." and ``Quant. Corr.", respectively.
\label{tab:tc1xxx1}}
\end{center}
\end{table}
\begin{table}[h]
\begin{center}
\begin{tabular}{cccc}
\hline Condition& Strategy&$$ & Payoff\\
\hline
&&&\\
 {\bf No Corr.}& No NE&&---   \\
 &(\it{pure strategy})&&   \\
&&&\\
 &$(\frac{1}{2}, \frac{1}{5})$&&$(-\frac{1}{5},\frac{3}{2})$ \\
 &&&\\

{\bf Quant. Corr.}&\hspace{5mm}No NE&&---\\
&(\it{pure strategy})&&   \\
\\
 $\frac{|00\rangle+i|11\rangle}{\sqrt{2}}$&$(\frac{1}{2},\frac{1}{5})$&&$(-\frac{1}{5},\frac{3}{2})$  \\
 \\
 $\frac{|01\rangle-i|10\rangle}{\sqrt{2}}$&$(\frac{1}{2}, \frac{4}{5})$&&$(-\frac{1}{5},\frac{3}{2})$  \\
 &&\\{\bf Class. Corr.}&No NE&&---\\
&~~(\it{pure strategy})&&   \\
\\
 $\frac{|00\rangle\langle00|+|11\rangle\langle11|}{2}$&$(\frac{1}{2}, \frac{1}{2})$&&$(\frac{1}{4},\frac{3}{2})$  \\
 \\
 $\frac{|01\rangle\langle01|+|10\rangle\langle10|}{2}$&$(\frac{1}{2}, \frac{1}{2})$&&$(\frac{1}{4},\frac{3}{2})$  \\
\\
 {\it Full rank} &$(\forall p, \forall q)$&&$(\frac{1}{4},\frac{3}{2})$  \\
\\
\hline
\end{tabular}
\caption{Effect of restricting the operations of the players to
the classical operators $\sigma_{0}$ and $i\sigma_{y}$ on the
strategies of the players to achieve NE. In pure strategies there
is no NE, no matter what kind of correlated state is shared
between the players. NE's appear only in mixed strategies $(p,q)$.
Payoffs $(\$_{A},\$_{B})$ and mixed classical strategies $(p,q)$
are given only for the NE points. $p$ and $q$ denote the
probabilities that the players apply $\sigma_{0}$. Classical and
quantum correlations are denoted by ``Class. Corr." and ``Quant.
Corr.", respectively. \label{tab:tc1yyy1}}
\end{center}
\end{table}
\noindent used $p=p_{0}$. The average payoffs can be calculated
from Eq. (\ref{N03}) as
\begin{eqnarray}\label{N14}
\hspace{-5mm}&&\$_{A}=\sum_{n} a_{n}P_{j \ell}=\sum_{n} a_{n}(pP_{j \ell}^{\prime}+(1-p)P_{j \ell}^{\prime\prime})\nonumber\\
\hspace{-5mm}&&\$_{B}=\sum_{n} a_{n}P_{j \ell}=\sum_{n}
b_{n}(pP_{j \ell}^{\prime}+(1-p)P_{j \ell}^{\prime\prime})
\end{eqnarray} from which we can write
\begin{eqnarray}\label{N15}
\hspace{-5mm}P_{00}&=&p\cos^{2}(\frac{\theta_{A}}{2})\cos^{2}\phi_{A}+(1-p)\sin^{2}(\frac{\theta_{A}}{2})\sin^{2}\varphi_{A}\nonumber\\
\hspace{-5mm}P_{01}&=&(1-p)\cos^{2}(\frac{\theta_{A}}{2})\cos^{2}\phi_{A}+p\sin^{2}(\frac{\theta_{A}}{2})\sin^{2}\varphi_{A}\nonumber\\
\hspace{-5mm}P_{10}&=&(1-p)\cos^{2}(\frac{\theta_{A}}{2})\sin^{2}\phi_{A}+p\sin^{2}(\frac{\theta_{A}}{2})\cos^{2}\varphi_{A}
\end{eqnarray}

When $\hat{U}_{A}$ is chosen from one-parameter SU(2) space that
is $\phi_{A}=\varphi_{A}=0$, there is an NE for $p=0.2$ and
$\theta_{A}=\pi/2$ with Alice and Bob having payoffs $-0.2$ and
$1.5$, respectively. This payoff is the same as the case when both
players are restricted to use a quantum pure strategy with
one-parameter SU(2) operators which also gives the same result of
classical mixed strategies. Although this strategy introduces an
NE into the game on which both players can compromise, it is not
the most desirable case for Alice due to her negative payoff while
Bob's is positive (CASE I). Indeed, when the payoffs are analyzed
in detail, it can be seen that Alice can never achieve, for any
value of $\theta_{A}$ a positive payoff independent of Bob's
strategy defined by $p$.

If Alice chooses her operator from two-parameter SU(2) operators,
then it is seen that there is no NE in the game and there is no
strategy for Alice to have a positive payoff. For any strategy of
Alice, Bob can find a $p$ value that will minimize Alice's payoff
and vice verse.

When Alice decides to choose her operator as a general SU(2)
operator, she will have three parameters to optimize her payoff.
In this case, we find that payoffs of both players become
independent of Bob's strategy when Alice's strategy is either
$(\theta=\pi/2,\phi=\varphi=\pi/4)$ resulting in
$\hat{U}_{A}=[\sigma_{0}+i(\sigma_{x}+\sigma_{y}+\sigma_{z})]/2$
or $(\theta=\pi/2,\phi=0,\varphi=\pi/2)$ resulting in
$\hat{U}_{A}=(\sigma_{0}+i\sigma_{x})/\sqrt{2}$. The players get
the payoffs $(0.25,1.5)$ and $(1,2.5)$, respectively, for the
first and second strategies (CASE II). On the other hand, if Alice
applies the operator $(\theta=\pi/2,\phi=0,\varphi=\pi/4)$ that is
$\hat{U}_{A}=[\sigma_{0}+i(\sigma_{x}+\sigma_{y})/\sqrt{2}~]/\sqrt{2}$
or $(\theta=\pi/2,\phi=\pi/4,\varphi=\pi/2)$ that is
$\hat{U}_{A}=[\sigma_{0}+i(\sqrt{2}\sigma_{x}+\sigma_{z})]/2$,
then while Bob's payoff becomes 2 and is independent of his own
strategy, the payoff of Alice is dependent on Bob's strategy
through $\$_{A}=(1+3p)/4$ and $\$_{A}=1-3p/4$, respectively, for
the first and second operators. With these moves of Alice, both
players get positive payoffs which resolves the dilemma in the
weak sense (CASE II).

A selfish and evil-thinking Bob might try to prevent Alice having
a positive payoff. Then the only thing he can do is to induce
damping on his qubit until the initial MES becomes a mixed state
with no off-diagonal elements (classical correlation). However, in
this case, Alice can choose her operator such that
$\theta_{A}=\pi/2$ making both her and Bob's payoff independent of
Bob's strategy. This results in a positive payoff for both players
with $\$_{A}=0.25$ and $\$_{B}=1.5$ resolving Alice's dilemma as
in CASE II.

It is understood that by acting unilaterally Alice can resolve her
dilemma only when she is allowed to use operators chosen from
three-parameter SU(2) set and Bob is restricted to classical
operations. Even in this case, only a solution satisfying CASE II
is achieved. When Bob is also allowed to use the same type of
operator, for each operator of Alice, Bob can always find an
operator which will put Alice into dilemma.

\section{V. CONCLUSIONS and discussions}

In this paper, we have discussed the effects of classical and
quantum correlations in the Welfare Game using the quantization
scheme proposed by Eisert et al. \cite{Eisert1}. This comparative
study of the effects of classical/quantum correlations and
classical/quantum operations not only confirmed some of the known
effects of using the quantum mechanical toolbox in game theory but
also revealed that in some circumstances one need not stick to
quantum operations or correlations but rather use simple classical
correlations and operations. This can be clearly seen in Tables
\ref{tab:tc1xxx1}-\ref{tab:tc1xz} which summarize the results of
this study.

The significant effect of using the quantum operations with shared
quantum correlation between the players is the emergence of NE's
which turn the game into a coordination game. Together with the
results of other games studied using the quantum mechanical
toolbox, we can say that there is at least one NE in $2\times2$
games with arbitrary payoff matrices when played using one or two
parameter SU(2) operators. Sometimes there can emerge more than
one NE with either the same or different payoffs based on the
structure of the payoff matrix of the original game. When this
happens, players are indifferent between the multiple NE. Then it
becomes unclear how the players
\begin{table}[h]
\begin{center}
\begin{tabular}{cccc}
\hline ~~~~~~~~~~&Class. Op.$~~~$&\multicolumn{2}{c}Quant. Op.\\
\hline~~~~~~~~~~~&&$~~{\rm Q_{1}}$&${\rm Q_{2}}$\\ \hline
\hspace{-16mm}\textbf{No. Corr.}&I&I& I \\
\hspace{-12mm}\textbf{Class. Corr.}&&\\
\hspace{6mm}$(|00\rangle\langle00|+|11\rangle\langle11|)/2$&II&II&II \\
\hspace{6mm}$(|01\rangle\langle01|+|10\rangle\langle10|)/2$&II&II&III\\
\emph{Full rank}&II&II&II\\
\hspace{-10mm}\textbf{Quant. Corr.}&~~&\\
$\hspace{6mm}(|00\rangle+i|11\rangle)/\sqrt{2}$&I&I&
III\\
$\hspace{6mm}(|01\rangle-i|10\rangle)/\sqrt{2}$&I&I&
n.a\\
\hline
\end{tabular}
\caption{Solutions to the dilemma in the Welfare game when the
players choose classical and quantum strategies. I, II and III,
respectively, represent CASE I: $\$_{A}<0$, CASE II:
$0\leq\$_{A}\leq\$_{B}$, and CASE III: $0\leq\$_{B}<\$_{A}$. n.a
which stands for ``not applicable" points out the absence of a
unique NE in the game. For the classical operations, only mixed
strategies are listed because in pure strategies there is no
unique NE in the game. In case of quantum operations, ${\rm
Q_{1}}$ and ${\rm Q_{2}}$ stand for one- and two-parameter SU(2)
operators, respectively. \label{tab:tc1xz}}
\end{center}
\end{table}
should behave, therefore this NE cannot be taken as a solution for
the problem. In such cases, introducing the quantum mechanical
toolbox does not make the things easier but rather more
complicated.

The surprising result of this study is that the dilemma of Alice
can be resolved with the weak (CASE II) and strong (CASE III)
solutions even with classical correlations provided that the
players are allowed to use quantum operations. Another point to be
noted is that the strategies of the players to find an NE and the
payoffs they get at this NE depend on the type of the shared
quantum correlation or classical correlation. Without a priori
information on the distributed or shared correlations, the players
cannot make their move.

The power of entanglement for this specific game shows itself in
the amount of payoff players get but not whether it resolves the
dilemma or not. When MES are allowed with two-parameter SU(2)
operators, NE's with payoffs $(3,2)$, which has the highest
$\$_{A}+\$_{B}$ value achievable in this game, emerge. When the
dilemma is resolved with classical correlations, the payoffs the
players get are much smaller, $\$_{A}+\$_{B}=7/2$.

We have to point out that, introducing correlations into a
$2\times2$ noncooperative game transforms it into a cooperative
game which is very different than the original game defined with
its own rules. In the original classical game, no communication of
any form is allowed between the players; however in the full
quantum strategies, shared entanglement which could be considered
as a kind of {\it ``spooky communication"} between the players are
introduced into the game \cite{Text2}.

In classical mixed strategies, the payoffs become continuous in
the mixing probability of the players' actions, and a compromise
becomes possible between the players which assures the existence
of an NE. In games played by quantum operations, the inclusion of
new strategies and moves into the game is an essential feature.
Because, once these new moves are incorporated, both the strategy
space and the payoffs become continuous. With this continuity, the
players can finely adjust their strategies and come closer to
being best responses to each other which leads to one or more
NE's.

\begin{acknowledgments}
The authors thank  Dr. T. Yamamoto for his help and stimulating
discussions.
\end{acknowledgments}

\end{document}